\documentclass[12pt, a4paper]{iopart}

\usepackage{graphicx}
\usepackage{amssymb}
\usepackage{braket}
\usepackage{hyperref}
\usepackage{cite}
\usepackage{color}
\usepackage[normalem]{ulem}

\begin{document}

\title[]{Many-body localization, symmetry, and topology}

\author{S. A. Parameswaran$^{1}$ and Romain Vasseur$^{2}$}
\address{${}^1$Rudolf Peierls Centre for Theoretical Physics, University of Oxford, 1 Keble Road, Oxford OX1 3NP, UK}
\address{${}^2$Department of Physics, University of Massachusetts, Amherst, MA 01003, USA}

\ead{sid.parameswaran@physics.ox.ac.uk, rvasseur@umass.edu}
\vspace{10pt}

\begin{abstract}
We review recent developments in the study of out-of-equilibrium topological states of matter in isolated systems. The phenomenon of many-body localization, exhibited by some isolated systems usually in the presence of quenched disorder, prevents systems from equilibrating to a thermal state where the delicate quantum correlations necessary for topological order are {often} washed out. Instead, many-body localized systems can exhibit a type of {\it eigenstate phase structure} wherein their entire many-body spectrum is characterized by various types of quantum order, usually restricted to quantum ground states. After introducing many-body localization and  explaining how it can protect quantum order, we then explore how the interplay of symmetry and dimensionality with many-body localization constrains its role in stabilizing topological phases out of equilibrium. 
\end{abstract}


\section{Introduction}
A central focus of condensed matter physics is to identify and classify different phases of matter. Traditionally, this activity has focused on systems in thermal equilibrium. At zero temperature ($T=0$), distinct equilibrium phases of matter correspond to distinct quantum ground states. These may be classified in terms of symmetries, or as more recently recognized, via their topological properties. The latter realization has led to the identification of topological phases of matter not captured by the conventional Landau paradigm of symmetry breaking~\cite{wen2004quantum,Hasan2010,Moore:2010aa}. Topological phases are characterized by  the entanglement structure of their quantum wavefunctions, and often host exotic surface states, as exemplified by the celebrated topological insulators~\cite{Hasan2010,Moore:2010aa}.

At finite temperature, many-body systems often have a less rich phase structure in physically accessible dimensions. Thermal fluctuations, stronger in low dimensions, are particularly detrimental to the subtle quantum correlations responsible for topological order, but they also destabilize more conventional orders.  For instance, though (broken) symmetry continues to be relevant even for $T>0$, thermal equilibrium strongly constrains possible phases: as an example, fluctuations destroy symmetry-breaking in one spatial dimension ($d=1$).

Recently, it has been recognized that some isolated quantum systems {\it far} from equilibrium can have richer phase structure than previously thought possible. 
When the `resonances' between eigenstates that are responsible for thermalization are suppressed --- e.g., by the presence of quenched disorder --- isolated systems can exhibit quantum coherent phenomena without being cooled to low temperature, via the phenomenon of ``many-body localization'' (MBL)~\cite{FleishmanAnderson,Gornyi,BAA,PhysRevB.75.155111,PalHuse,HuseMBLQuantumOrder}. Due to the localization~\cite{PhysRev.109.1492} of their excitations --- even when a finite density of them is present ---  MBL systems cannot reach thermal equilibrium under their own dynamics~\cite{2014arXiv1404.0686N}: in other words, {they do not {\em thermalize}}. Instead, their highly-excited eigenstates are similar --- in a sense precisely quantifiable via entanglement --- to ground states~\cite{BauerNayak,PhysRevLett.111.127201}. 
 Distinct non-equilibrium phases may then be linked to the symmetry and/or topological properties of the eigenspectrum as a whole. The resulting {\it eigenstate phases}~\cite{HuseMBLQuantumOrder,BauerNayak,BahriMBLSPT,PhysRevB.89.144201} represent the natural generalization of $T=0$ quantum phases to finite energy density, the analog of ``high temperature'' for an isolated system  not in thermal equilibrium.  Such {\it localization protected quantum orders}~\cite{HuseMBLQuantumOrder} include topological phases and possibilities forbidden in equilibrium, and form the subject of this review.

After introducing key aspects of MBL and localization-protected symmetry breaking order, we explain how non-Abelian symmetries impact localization. Building on this we discuss two distinct classes of  non-equilibrium topological phases:
\begin{itemize}
\item {\it Non-equilibrium symmetry-protected  topological phases. } We explain how MBL can stabilize non-equilibrium  symmetry protected topological (SPT) phases (e.g., topological insulators), and by analyzing constraints imposed by symmetry and dimensionality, present 
a tentative classification of non-equilibrium SPTs.
\item {\it Non-equilibrium topologically ordered phases. } We then discuss how MBL can stabilize long-range entangled topologically ordered phases, and address how the ensuing localization of anyons impacts topological quantum computing.

\end{itemize}
We close with a summary of various subtleties elided in the bulk of this review, as well as some comments on potentially interesting future directions.

\section{Localization protected quantum order and global symmetries}

\subsection{A short review of many-body localization}
We begin with a {\it precis} of aspects of many-body localization relevant to our discussion. 
Consider the random-field XXZ spin chain, with Hamiltonian (here, $h_i\in [-W,W]$):
\begin{equation}
\label{eq:XXZHam}
H = -\sum_i J_{\perp} (\sigma^x_i \sigma^x_{i+1} + \sigma^y_i \sigma^y_{i+1})+ h_i\sigma^x_i + J_z\sigma^z_i\sigma^z_{i+1},
\end{equation}
which is related by a Jordan-Wigner mapping to a spinless fermion chain with hopping $J_\perp$, on-site potential $h_i$, and a nearest-neighbor density-density interaction $J_z$. $H$ has a  U(1) symmetry corresponding to conserved magnetization (particle number) in the spin  (fermion) language. For $J_z=0$, the non-interacting model is Anderson localized by the quenched randomness. Many-body localization (MBL) refers to the persistence of this localization with interactions, $J_z\neq 0$ --- a fact initially demonstrated using diagrammatic perturbation theory~\cite{BAA}, extensively tested numerically~\cite{PalHuse,PhysRevB.75.155111,Luitz,PhysRevLett.114.160401,PhysRevB.93.060201}, and finally proved rigorously for a slightly different model~\cite{2014arXiv1403.7837I,PhysRevLett.117.027201}. 

An MBL system is a generic (i.e., not fine-tuned) interacting system that does not thermalize under its own dynamics (see~\cite{2014arXiv1404.0686N,doi:10.1146/annurev-conmatphys-031214-014701,1742-5468-2016-6-064010,ANDP:ANDP201700169,2017arXiv171103145A} for recent reviews). 
For our example, this is particularly clear in a phenomenological description that exposes its `emergent integrability', a hallmark of MBL. For $J_z=0$, it  the single-particle problem can be straightforwardly diagonalized. In the fermionic language, many-body eigenstates are then Slater determinants of orthogonal, exponentially localized single-particle orbitals with energy $\epsilon_i$. This means that there is a unitary transformation $\mathcal{U}$ such that (in the spin language) $ H = \sum_i \epsilon_i \tau^z_i$, where  
$\tau^\mu_i = \mathcal{U} \sigma_i^\mu \mathcal{U}^\dagger =\sum_{j,\nu}f^{\mu\nu}_{ij} \sigma^\nu_j$ are a new set of Pauli operators. The non-interacting and localized nature of the model is reflected by the linearity of this transformation and the fact that $f^{\mu\nu}_{ij} {\sim}e^{-|i-j|/\xi}$ as $|i-j|\rightarrow \infty$ for some finite localization length $\xi$.  Crucially, the Hamiltonian is now written in terms of an extensive set of commuting operators $\tau^z_i$, where the single-particle fermionic orbitals have occupancy $n_i = \frac{1+\tau^z_i}{2}$.

In an MBL system,  a modification of this statement continues to hold  even in the interacting case~\cite{PhysRevLett.111.127201,PhysRevB.90.174202}. Namely, there is again a unitary rewriting of $H$ in terms of a set of commuting operators $\tau^z_i=\mathcal{U} \sigma_i^z \mathcal{U}^\dagger$; however, these are now {\it nonlinear} combinations of  physical spins, and the Hamiltonian now has  higher-order couplings between the $\tau^z_i$s:
\begin{equation}
\label{eq:XXZlbitHam}
\!\!\!\!\!\!\!\tau^\mu_i = \sum_{j,\nu}f^{\mu\nu}_{ij} \sigma^\nu_j + \sum_{j,k,\nu,\rho} g^{\mu\nu\rho}_{ijk}\sigma^\nu_j\sigma^\rho_k +\ldots ,\,\,\,\,\,\, H = \sum_i \epsilon_i \tau^z_i + \sum_{ij}\mathcal{J}_{ij} \tau^z_i\tau^z_j +\ldots, 
 \end{equation}
where $g^{\mu\nu\rho}_{ijk}, \mathcal{J}_{ij} $ are $O(J_z)$ and decay exponentially in space, and the ellipsis denotes $(n\geq 3)$-body terms. The  $\tau^z_i$s are {\it localized integrals of motion} (LIOMs) or `l-bits' that commute with $H$, unlike the physical $\sigma^z_i$ `p-bits'~\cite{PhysRevLett.111.127201,PhysRevB.90.174202,Ros2015420,2014arXiv1403.7837I,PhysRevLett.117.027201}. A system with an {\it extensive} set of such LIOMs cannot thermalize. Unlike traditional integrable systems, here generic small perturbations that seemingly destroy integrability simple lead to a  redefinition of the l-bits. 
This phenomenological l-bit description, assumed throughout this review, captures many features of MBL, such as area-law entanglement~\cite{BauerNayak,PhysRevLett.111.127201} and logarithmic entanglement growth~\cite{PhysRevB.77.064426,PhysRevLett.109.017202}/dephasing~\cite{BahriMBLSPT,PhysRevLett.113.147204,PhysRevB.91.140202,PhysRevB.90.174302,PhysRevB.95.060201,2016arXiv160802765C,ANDP:ANDP201600318} after a quench from an initial product state.  {We will discuss below how this specific definition of MBL is not compatible with certain types of quantum order.} 

\subsection{Dynamical phase diagram of the random Ising chain}
We next consider the random Ising chain, given by the Hamiltonian
\begin{equation}
\label{eq:RTFIMHam}
H = -\sum_i J_i \sigma^z_i \sigma^z_{i+1} + h_i\sigma^x_i + J_i' \left(\sigma^z_i\sigma^z_{i+2} +\sigma^x_i \sigma^x_{i+1}\right)  .
\end{equation}
where the couplings are all random positive numbers, drawn from a distribution of width $W$. For $J'_i =0$ the model can be mapped to a spinless p-wave superconductor (sometimes called the Kitaev chain). The $J'_i \neq 0$ terms introduce interactions thus destroying free-fermion integrability, but preserve the self-duality of the Ising model.

In the clean case, $J_i = J$, $h_i = h$ and $J_i'=J'$, there is a $T=0$ phase transition between  a $J\gg h$ ferromagnet (FM) with a pair of degenerate ground states and  $J\ll h$ paramagnet (PM) with a unique ground state. These correspond, in the fermionic language, to topological (trivial) phases with (without) zero-energy Majorana end states. For $T>0$, we must consider excitations about these states. In the FM, these are domain walls between the two degenerate ground states. When the system is translationally invariant the domain walls are mobile  and can freely move across the sample, destroying the order. 
Thus at finite temperature the $d=1$ clean Ising model has a crossover rather than a  transition as $h/J$ is tuned. 

\begin{figure}[t]
\begin{center}
\includegraphics[width=\columnwidth]{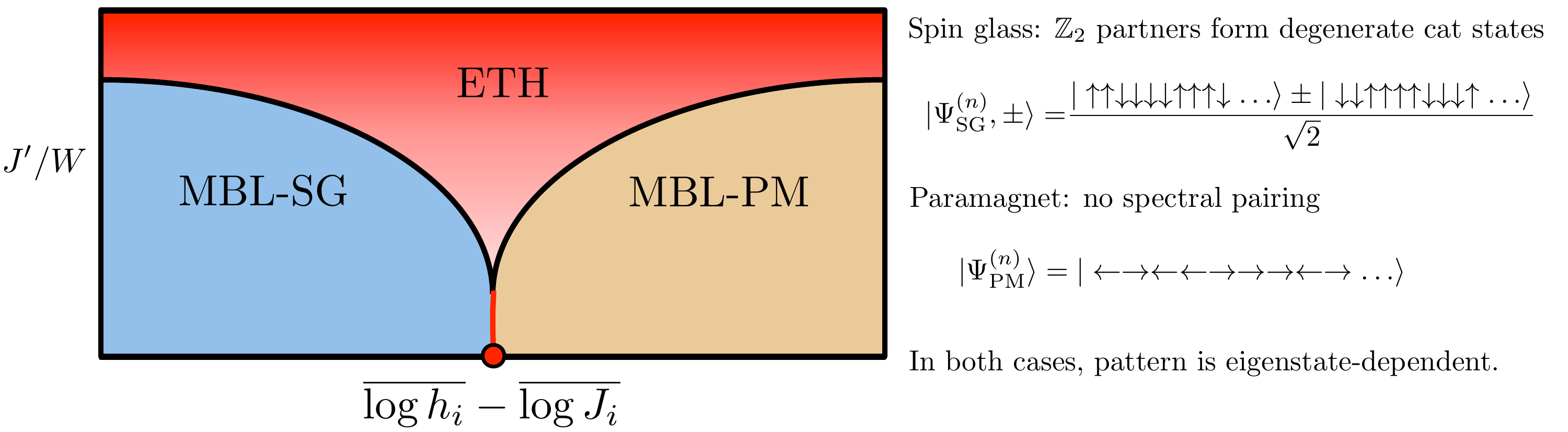}
\end{center}
 \caption{\label{fig:GlobalPD}Schematic non-equilibrium (eigenstate) phase structure of the random quantum Ising chain. Disorder is parametrized by $W$;  $h$ tunes the system across the transition. $J'$ is the interaction strength {which favors thermalization (ETH phase)}. {At strong disorder, two distinct MBL phases are possible.} See text for discussion.}
\end{figure}

For sufficiently strong disorder $W$ or sufficiently weak interactions $J_i'$, the system can be MBL. In this case, the domain walls of the FM are pinned by disorder, even when a finite density of them is present. In the Majorana language, this corresponds to the localization of a finite density of point particle excitations. Thus, with strong disorder, one can argue that  when the typical bond term dominates the typical field term, $\overline{\log J_i} \gg\overline{\log h_i}$, 
 the system exhibits
`spin glass order' (SG)~\cite{HuseMBLQuantumOrder,PekkerRSRGX}, in which $\langle \sigma^z_i \sigma^z_j\rangle$ has a nonzero expectation value even for $|i-j|\rightarrow\infty$, though the sign fluctuates with $i-j$ since the spins are `frozen' into some eigenstate-dependent configuration (unlike a ferromagnet, they do not all point in the same direction). In the opposite limit of very strong transverse fields ($\overline{\log J_i} \ll\overline{\log h_i}  $), the system is still MBL for strong disorder, but the spin-glass order is lost: this corresponds to an MBL {\it paramagnet}. So, the distinction between PM/SG persists across the entire eigenspectrum for strong enough disorder. For weaker disorder (stronger interactions) MBL is destroyed in favor of a thermal phase where there is no distinction between PM and SG: hence the SG phase is an instance of `localization-protected order'. The existence of  MBL PM/SG phases has been numerically verified~\cite{PhysRevLett.113.107204}. A global eigenstate phase diagram for the self-dual case is shown in  Fig.~\ref{fig:GlobalPD}. Note that whether the critical point between MBL PM/SG phases can itself be athermal for very strong disorder~\cite{PekkerRSRGX,VoskAltmanPRL13,PhysRevLett.112.217204,QCGPRL,PhysRevB.93.104205,1742-5468-2016-3-033101,ANDP:ANDP201600302}, or if instead a sliver of intervening thermal phase persists except at infinite disorder remains a matter of debate~\cite{2017arXiv171100020V,PC_KH,PC_SS}. Nevertheless there are compelling arguments that the critical point/crossover (as the case may be) between the MBL PM/SG is governed by an infinite-randomness fixed point accessed using a strong-disorder real-space renormalization group approach~\cite{PhysRevB.51.6411}.

\subsection{Localization-protected orders and non-Abelian symmetries}
As we have seen above,  MBL admits the possibility of realizing quantum orders far from thermal equilibrium, even in situations where they would be forbidden in equilibrium, e.g. by Mermin-Wagner or Peierls-Landau-type  arguments. This immediately raises the question: can all equilibrium phases be localization-protected  in the non-equilibrium setting? Perhaps more interestingly, can MBL stabilize non-equilibrium phases with no equilibrium counterpart? 
In fact, MBL imposes a relatively rigid structure (i.e., the l-bit description) which is incompatible with many types of order. A simple model, believed to evade MBL even for strong disorder, is the random-bond spin-$\frac{1}{2}$ Heisenberg chain
\begin{equation}
H = \sum_i J_i \vec{S}_i\cdot \vec{S}_{i+1},
\label{eqXXX}
\end{equation}
with random couplings $J_i$. 
First,  the  $SU(2)$ symmetry of~(\ref{eqXXX}) is incompatible with the area-law structure of MBL systems~\cite{PhysRevB.94.224206,PhysRevB.96.041122}. Moreover, renormalization-group~\cite{PhysRevB.89.144201,QCGPRL} and resonance-counting arguments~\cite{PhysRevB.96.041122} indicate that a non-ergodic phase with a more exotic (logarithmic) entanglement structure is unstable to thermalization. These results indicate that no matter how strongly disordered eq.~(\ref{eqXXX}) is, this system remains thermal (``ergodic''), though the dynamics {\it en route} to thermalization may well be glassy~\cite{PhysRevB.92.184203,PhysRevB.96.041122}.





The crucial feature of~(\ref{eqXXX}) that prevents MBL is its non-abelian 
$SU(2)$ symmetry group. In fact, even discrete non-abelian groups -- as opposed to Lie groups like $SU(2)$ -- suffice to forbid symmetry-preserving MBL phases. 
(This is  particularly salient where the symmetry is necessary to protect topological features, as we discuss below.) 
For instance, consider generalizing (\ref{eq:RTFIMHam}) to the quantum Potts chain, 
%
\begin{equation}
H =  -\sum_j J_j {\sigma}^{\dagger}_j {\sigma}^{\dagger}_{j+1} - h_j {\tau}^{\dagger}_j    + {\rm h.c.},
\label{eqPotts}
\end{equation}
where the operators $\sigma_j$ and $\tau_j$ act on a three-dimensional single-site Hilbert space via $\sigma \ket{m} = {\rm e}^{\frac{2 \pi m i}3} \ket{m}$ and ${\hat{\tau}} \ket{m} = \ket{m+1}$, with the ket labels taken modulo $3$.  This Hamiltonian has  a global ${\mathbb{Z}_3}$ rotation symmetry generated by ${\hat{\mathcal{Q}}} = \prod_j \tau_j$ and a {$\mathbb{Z}_2$ mirror symmetry ${\hat{\mathcal{X}}} \equiv \prod_j \mathcal{X}_j$, where $\mathcal{X}_j$ exchanges the $\ket{1}$, $\ket{2}$ eigenstates of $\hat{\sigma}_j$; clearly,   ${\hat{\mathcal{Q}}}^3={\hat{\mathcal{X}}}^2 =1$. Together, ${\hat{\mathcal{Q}}}, {\hat{\mathcal{X}}}$ generate the permutation group $S_3\cong {\mathbb{Z}_3}\rtimes {\mathbb{Z}_2}$, where the semidirect product structure reflects the fact that ${\hat{\mathcal{Q}}}$,  ${\hat{\mathcal{X}}}$ do not commute. This model was studied numerically in Refs.~\cite{2017arXiv170600022F,PhysRevB.96.165136}. Unlike ~(\ref{eqXXX}), model~(\ref{eqPotts}) supports two distinct broken-symmetry MBL phases at strong disorder, that correspond to breaking either the ${\mathbb{Z}_3}$ clock or the ${\mathbb{Z}_2}$ chiral symmetries. However, in contrast to the Ising case, there is {\it no} symmetry-preserving MBL paramagnet:  MBL phases necessarily  break the $S_3$ non-abelian symmetry spontaneously down to an Abelian subgroup. The  Heisenberg and Potts results are examples of a general ``no-go'' theorem discussed below.

\section{Non-equilibrium symmetry protected topological phases} 

Following the theoretical prediction and subsequent experimental discovery of topological insulators and superconductors~\cite{PhysRevLett.95.226801,Bernevig1757,Konig766,PhysRevLett.98.106803,PhysRevB.75.121306,PhysRevB.79.195322,Hsieh:2008aa,Hasan2010,Moore:2010aa}, several different short-range entangled phases with topological edge modes protected by symmetry have been identified~\cite{PhysRevB.80.155131,PhysRevB.84.235128,PhysRevB.83.075102,PhysRevB.83.075103,Chen2011b,Pollmann2012,YuanMing2012,Levin2012,Chen1604,Chen2011,doi:10.1063/1.3149495,PhysRevB.78.195125,PhysRevB.81.134509,PhysRevB.89.201113,Wang629,PhysRevB.90.115141,Gu2015}. These 
SPT phases may also emerge in strongly interacting systems, including for instance the celebrated Haldane phase in quantum spin chains~\cite{Buyers1986}. 
There is now an exhaustive classification of gapped bosonic~\cite{Chen1604,Chen2011,YuanMing2012} and to some extent, fermionic, SPT phases~\cite{PhysRevB.55.1142,doi:10.1063/1.3149495,PhysRevB.78.195125,PhysRevB.81.134509,PhysRevB.83.075103,PhysRevB.89.201113,Wang629,PhysRevB.90.115141,Gu2015,PhysRevLett.117.206405}. 
MBL could in principle be used to protect topological edge modes far from equilibrium, without the need for cooling~\cite{BahriMBLSPT,PhysRevB.89.144201}. However, this exciting possibility is tempered by various constraints on the existence of MBL, reviewed below, 
 which rule out the existence of many MBL SPT phases, including, for example, electronic topological insulators protected by time-reversal symmetry. 

\subsection{Chiral edge modes}
In two dimensions there exist chiral topological phases, such as those exhibiting the integer quantum Hall effect, that neither require protecting symmetries nor host fractionalized bulk excitations.
 There is a fundamental obstruction 
  to constructing a commuting projector  (``zero correlation length'') Hamiltonian~\cite{KITAEV20062,PhysRevB.89.195130}  for phases with chiral edge modes. 
  Intuitively, this arises as a contradiction: edge states of chiral phases necessarily carry some heat current (there is a `thermal Hall effect'), yet this must vanish identically 
  for commuting projector Hamiltonians. 
  As the l-bit structure of MBL necessarily implies the existence of a commuting-projector limit, this  
  demonstrates that chiral phases cannot be many-body localized~\cite{NandkishorePotterScaling,2015arXiv150600592P}. Interestingly,  this obstruction can be avoided 
  in periodically-driven (Floquet) systems whose energy is not conserved~\cite{PhysRevX.6.041070,PhysRevLett.118.115301}. 
  

\subsection{Particle-hole and time-reversal symmetries}

Many 
 SPT phases are protected by time-reversal and/or particle-hole symmetries, which can be also be an obstruction to MBL. 
  In a system where time-reversal symmetry (TRS) is represented by an anti-unitary operator ${\cal T}$ that satisfies ${\cal T}^2 = \pm 1$, eigenstates 
  any eigenstate $\Ket{\Psi}$ is always degenerate with its Kramers partner ${\cal T} \Ket{\Psi}$ if ${\cal T}^2=-1$.
   In an MBL system with a commuting-projector limit, TRS implies that the l-bits also label {\em local} Kramers doublets~\cite{PhysRevB.94.224206} (we give a general argument in the next section). These local degeneracies in turn imply an exponential-in-system-size degeneracy of the many-body eigenstates, that cannot be stable under generic perturbations. Importantly, these local degeneracies can be lifted only by breaking the symmetry (spontaneously or explicitly), or by giving up the local integrability structure currently thought to be a defining feature of MBL. 
This argument implies that time-reversal protected topological phases with ${\cal T}^2=-1$ cannot be protected by MBL.  
This means that it is impossible to localize time-reversal invariant systems of spin-$\frac{1}{2}$ electrons, ruling out the use of MBL to stabilize electronic topological insulators out of equlibrium.

 A similar argument can be made for particle-hole symmetric (PHS) systems with symmetry group $U(1) \rtimes {\mathbb Z}_2$. Here, the $U(1)$ subgroup corresponds to particle number conservation, and the semidirect product  (`$\rtimes$') structure arises from the fact that $U(1)$ representations are exchanged by the particle-hole ${\mathbb Z}_2$ symmetry. The non-Abelian nature of $U(1) \rtimes {\mathbb Z}_2$ causes two-fold degeneracies~\cite{PhysRevB.93.134207} similar to the Kramers degeneracies discussed above, preventing a particle-hole preserving MBL phase by the same argument.  Note that it {\it is} possible to realize an MBL phase if PHS is broken spontaneously (in a spin-glass fashion as in the Ising example). However,  the resulting phase could not host topological edge modes if PHS is part of the protecting symmetry group of the SPT~\cite{PhysRevB.93.134207}.

\subsection{General symmetry constraints on many-body localization}

The examples above motivate the conclusion that
MBL is impossible in any system whose global symmetry group has multiplets (i.e. irreducible representations or irreps with dimension $>1$)~\cite{PhysRevB.94.224206}. This includes all non-Abelian symmetry groups, either continuous or finite. Intuitively, this is because the l-bits that define MBL systems transform independently under the global symmetry, leading to an extensive number of local degeneracies in the presence of non-Abelian symmetries. More precisely, if we were to have an MBL system with non-Abelian symmetry $G$, then the l-bits would transform as irreps of the symmetry group $G$ so that different states of each l-bit label irreps of $G$. If $G$ is non-Abelian and acts faithfully on the system, some of these irreps must have dimension larger than one leading to local degeneracies. These  in turn lead to an exponential-in-system-size degeneracy of all eigenstates. Such degenerate eigenstates are inherently unstable to perturbations, and there is no local and symmetry-preserving way to lift them. Hence, either symmetry or localization must break down. This argument forbids MBL-protected SPT phases with non-Abelian symmetry groups~\cite{PhysRevB.94.224206}. (For these purposes, TRS with ${\cal T}^2=-1$ can be considered as non-Abelian.) 

\subsection{Dimensionality dependence of many-body localization}

In many symmetry classes, nontrivial SPT states only exist in spatial dimension $d\geq 1$. So, in addition to symmetry constraints, an added complication is the dimensional-dependence of the stability of MBL. This is argued to be strong~\cite{PhysRevB.95.155129} (see also Refs.~\cite{Ponte20160428,ANDP:ANDP201600326}), by an argument whose simplified version runs as follows: consider 
a rare thermal region of radius $r_0$ in an otherwise localized system. 
Such a rare thermal inclusion, unavoidably in a generic disordered system, will act as a bath and will thermalize peripheral spins, increasing its radius to $r_0 + r$. Assuming the thermal ``bubble'' of radius $R=r_0 + r$ is described by random matrix theory (and thus remains featureless after absorbing many l-bits), the matrix element for flipping an l-bit at distance $R$ from the center of the inclusion is given by $\sim {\rm e}^{-r/\xi_\star}/\sqrt{{\rm dim} {\cal H}_R}$ where ${\rm dim} {\cal H}_R \sim {\rm e}^{C R^d}$ is the dimension of the Hilbert space of the bubble (initial thermal inclusion + closer l-bits). This matrix element should be compared to the renormalized level spacing of the bubble $\delta \sim 1/{\rm dim} {\cal H}_R$.  In one dimension if $\xi_\star$ is small enough (strong disorder), the matrix element falls off faster with distance $r$, and this ``avalanche'' process eventually stops so the inclusion is unable to thermalize far enough l-bits. For $d>1$ however, a sufficiently large thermal inclusion ($r_0$ large enough) appears sufficient to destabilize the MBL phase. This simple argument rests on some  dubious assumptions (in particular, it is unclear whether random-matrix theory still describes the bubble once it has absorbed a number of l-bits comparable to the number of spins in the initial inclusion of radius $r_0$~\cite{PhysRevB.95.134302,ANDP:ANDP201600326,EhudPrivateCom}), {some numerical studies are consistent with this scenario}~\cite{PhysRevLett.119.150602,Ponte20160428} and it does indicate that MBL might not be stable against thermalization in dimension $d>1$.  Note, however, that the extremely long thermalization timescales involved might make this instability an academic issue. Rare thermal regions can also be avoided altogether by considering deterministic quasi-periodic systems that show MBL~\cite{PhysRevB.87.134202,Schreiber842}. This underscores the fact that the fundamental requirement for a stable  MBL phase is not randomness, but instead a suitable `detuning' of resonances responsible for thermalization, whether achieved  via quenched disorder or other means, such as quasiperiodicity.

\subsection{Classification of non-equilibrium symmetry protected topological phases}

Gathering these different no-go results, we can now formulate a tentative classification of non-equilibrium SPT phases protected by MBL. We focus here on whether equilibrium SPT phases can be stabilized at finite energy density by MBL, though we briefly mention analogs in periodically driven (Floquet) systems in the discussion. 
 To summarize, starting from the equilibrium classification of interacting bosonic and fermionic SPT, a given topological phase is ``many-body localizable'' if~\footnote{Assuming that the constraints described above are exhaustive.}: (i) it is non-chiral, (ii) the protecting symmetry group does not protect degeneracies, (iii) it is one-dimensional, (iv) it has a well-defined, non-degenerate commuting-projector Hamiltonian limit~\cite{2015arXiv150600592P} (see also~\cite{CenkeMBLSPT}). Note that point (iv) includes (i) and (ii), and is in principle more general. Even if we ignore the potential instability of MBL in dimension $d>1$ ({{\it e.g.} by considering quasiperiodic systems}), constraints (i) and (ii) alone are sufficient to rule out the localizability of many fermionic SPTs, including the familiar electronic topological insulators (protected by charge conservation and time reversal symmetry with ${\cal T}^2=-1$), the Haldane chain with $SO(3)$ symmetry, or integer quantum Hall states. In addition to these constraints, we remark that  topological superconducting phases compatible with the above constraints may face additional practical difficulties as they require a pair condensate, which in ultra-cold neutral atom systems where MBL may be realized~\cite{Schreiber842,Smith:2016aa,PhysRevX.7.011034,Choi1547,2016arXiv161205249W}, implies the existence of a superfluid Goldstone mode with a diverging single-particle localization length at low energies, also believed  to be an obstacle to MBL~\cite{PhysRevB.27.5592,PhysRevB.68.134207,NandkishorePotterScaling,PhysRevLett.116.116601}.

\section{Non-equilibrium topologically ordered phases}
\label{sec:LPTO}
If we ignore for the moment the potential instability of MBL in dimension $d>1$~\cite{PhysRevB.95.155129} described above, MBL can also be used to protect topologically ordered phases with anyonic bulk excitations and long-range entanglement, as first recognized in Refs.~\cite{HuseMBLQuantumOrder,BauerNayak}. Such non-equilibrium phases with intrinsic topological order naturally have a finite density of localized anyonic excitations that could have interesting properties for topological quantum computing~\cite{RevModPhys.80.1083}, without the need of cooling below a spectral gap.

\subsection{Disordered toric code/$\mathbb{Z}_2$ gauge theory}
The simplest model that realizes a topologically ordered phase in two dimensions is Kitaev's toric code~\cite{KITAEV20032,2009arXiv0904.2771K}, described on the square lattice by the Hamiltonian
\begin{equation}\label{eq:HTC}
H_{\rm{TC}} = - \sum_{s} \Gamma^M_s A_s - \sum_p K_p B_p, \,\,\,\,{\rm with}\,\,\,\, A_s = \prod_{\ell: s \in \partial \ell} \sigma^x_{\ell}, \,\,\, B_p = \prod_{\ell \in \partial p} \sigma^z_{\ell},
\end{equation}
where $\ell, s, p$ label links, sites, and plaquettes,  and $A_s$ ($B_p$) acts on the four links that touch a site (encircle a plaquette) and the couplings are all positive. It is clear that $[A_s, B_p]=0$ and each term squares to the identity, so $H_{\rm{TC}}$ is of commuting-projector form. The ground state has $A_s = B_p =1$ and the elementary excitations are `electric charges' ($e$) and `magnetic fluxes' ($m$) which respectively live on sites with $A_s=-1$ and plaquettes with $B_p=-1$. These excitations have no dynamics in the toric code, but have a mutual statistical angle of $\pi$: encircling an $e$ with an $m$ or vice versa leads to a relative sign of $-1$ in the wavefunction.  On a torus, $H_{\rm{TC}}$ has a four-fold ground state degeneracy as the system size $L\rightarrow \infty$. This topological degeneracy is linked to the existence of  operators $\mathcal{X}_{1,2} =  \prod_{\ell \in  \mathcal{C}^{1,2}_s}\sigma^x_\ell$, where $\mathcal{C}^{1,2}_s$ are closed contours of sites that each encircle one of the two non-contractible loops of the torus and the product is over links contained in such a contour. Defining another pair of operators  $\mathcal{Z}_{1,2} =  \prod_{\ell \in  \mathcal{C}^{1,2}_p}\sigma^z_\ell$ we can show easily that $[H, \mathcal{X}_i ] = [H, \mathcal{Z}_i] =0$, $[\mathcal{X}_i, \mathcal{Z}_i] = 0$ but that $\mathcal{X}_1 \mathcal{Z}_2 = - \mathcal{Z}_2 \mathcal{X}_1, \mathcal{X}_2 \mathcal{Z}_1 = - \mathcal{Z}_1 \mathcal{X}_2$. From this it is clear that given an eigenstate $\Ket{\Psi}$ of $H$ which has has $\mathcal{Z}_1$, $\mathcal{Z}_2$ eigenvalues $z_1, z_2$,   we can construct three other eigenstates $\mathcal{X}_{1}\Ket{\Psi}, \mathcal{X}_{2}\Ket{\Psi}, \mathcal{X}_{1}\mathcal{X}_{2}\Ket{\Psi}$ with $\mathcal{Z}_1$, $\mathcal{Z}_2$ eigenvalues $(z_1, -z_2)$, $(-z_1, z_2)$, and $(-z_1, -z_2)$, respectively. This statement is true for {\it any} eigenstate in the commuting-projector limit, but in the clean limit where $K_p=K,  \Gamma^M_s = \Gamma^M$, perturbations away from this limit will spoil this for generic highly excited states. However, when the couplings are all random the four-fold degeneracy can be argued to be perturbatively stable in {\it every} eigenstate. 

To understand this physically, note that the quadruple degeneracy reflects the deconfinement of an electric charge excitation. Beginning in a given eigenstate, we can create an pair of $\mathbb{Z}_2$ electric charges from the vacuum, drag them around a noncontractible loop $\mathcal{C}_s$ of the torus, and then annihilate them (implemented by $\mathcal{X}_{1,2}$), leaving behind a string of flipped bonds connecting the sits in $\mathcal{C}_s$. No local perturbation can then distinguish the distinct eigenstates, but owing to the mutual statistics of $e/m$ encircling the `gauge string' with a magnetic excitation (implemented by $\mathcal{Z}_{1,2}$) can detect the presence/absence of a charge string. 
Thus, the four distinct ground states  of (\ref{eq:HTC}) correspond to the four possible values of  $(\mathcal{Z}_{1},\mathcal{Z}_{2}$). 
 Perturbing away from solvability, it is straightforward to show that splittings between these states are $O(e^{-L})$ since connecting  two topologically degenerate states requires going to $L^{\rm th}$ order in perturbation theory. However, at finite energy density in the clean system, away from the solvable point there will be a finite density of mobile $\mathbb{Z}_2$ electric charges present in the system, and in the presence of translational invariance local perturbations mix distinct flux sectors at $O(1)$ and destroy spectral quadrupling. Localization prevents this by rendering the $\mathbb{Z}_2$ charges immobile~\cite{HuseMBLQuantumOrder,BauerNayak}. This strongly suppresses the matrix elements for the tunneling between degenerate eigenstates, leaving the quadruple degeneracy intact. The four  states  that differ by the presence/absence of the charge string encircling the non-contractible loops will have the same pattern of frozen matter excitations.


The toric code can also be related to the $\mathbb{Z}_2$ (Ising) lattice gauge theory with matter, 
\begin{equation}
 H_{\rm{IGT}} =  -\sum_p K_p \prod_{\ell \in \partial p} \sigma^z_{\ell}
- \sum_{s}\Gamma^M_s \tau^x_s - \sum_{\ell} \left(\Gamma_\ell \sigma^x_\ell + J_\ell \sigma^z_\ell \prod_{s\in \partial\ell} \tau^z_\ell \right).
\end{equation}
where matter fields $\tau^\mu_s$ are defined on sites, and states in the physical Hilbert space satisfy  the local constraint $G_s\Ket{\Psi} = \Ket{\Psi}$ with $G_s = \tau^x_s\prod_{\ell: s \in \partial \ell} \sigma^x_{\ell}$. Evidently, $H_{\rm{IGT}}$ reduces to $H_{\rm{TC}}$ on states satisfying the constraint when $\Gamma_\ell=J_\ell=0$. $J_\ell$ ($\Gamma_\ell$) perturb away from exact solvability, inducing hopping of charges (fluxes). When they are large we enter a Higgs (confined) phase where the charges (fluxes) are condensed. Critical properties of this transition can be studied via the duality between $H_{\rm{IGT}}$ with  $J_\ell = \Gamma^M_s =0$  and a $2+1$D quantum Ising model where $K_p$ ($\Gamma_\ell$) map to bond (field) terms. 

{Since spectral quadrupling involves a finite-size splitting $\sim {\rm e}^{-L/\xi}$ typically much larger than the many-body level spacing $\sim  {\rm e}^{-\alpha L^2}$ for a system of linear size $L$, it is challenging to detect numerically, and one must turn to other diagnostics. 
} Eigenstate entanglement is one such diagnostic: similarly to a $\mathbb{Z}_2$-topologically ordered ground state~\cite{PhysRevLett.96.110404,PhysRevLett.96.110405}, we expect that in a phase where MBL protects deconfinement, generic excited states will have area-law scaling of entanglement entropy with a constant topological contribution of $\ln 2$.  
An alternative approach is to consider the scaling with length of Wilson loop operators, essentially the $\mathcal{X}_i,\mathcal{Z}_i$ considered above but with the contours now no longer encircling the system. The relevant distinction is between perimeter/area-law scaling in the deconfined/confined phases. At finite energy density in the presence of dynamical matter,  the Wilson-loop diagnostic will fail on general grounds~\cite{HuseMBLQuantumOrder,BauerNayak}. However various {\it ratios} of Wilson loops, collectively termed Fredenhagen-Marcu order parameters, continue to diagnose deconfinement even in this case~\cite{1367-2630-13-2-025009,HuseMBLQuantumOrder}.


\subsection{Many-body localization of anyons}

Generic eigenstates of MBL topologically ordered systems in 2+1 dimensions have a finite density of exponentially localized anyons. If such anyons are non-Abelian ({\it i.e.} if  they obey non-Abelian braiding statistics), the position of the anyonic excitations in the system is not sufficient to characterize an eigenstate: one must also specify a {\it fusion tree} splitting the degeneracies of all eigenstates with the same spatial configuration of anyons. This 
 is incompatible with the traditional l-bit structure of MBL phases, reflecting the absence of  local tensor product structure of the topological Hilbert space of non-Abelian anyons. In fact, if non-Abelian topologically ordered systems in 2+1 dimensions could be MBL, the finite density of localized non-Abelian anyons in generic eigenstates would lead to an exponential-in-size degeneracy of the eigenstates, generalizing the symmetry arguments above (with the quantum dimension of the anyons playing the role of the dimension of the irreps~\cite{PhysRevB.94.224206}). This forbids MBL phases with non-Abelian topological order, even in the absence of symmetry. Moreover, the symmetry constraints described above also forbid the localizability of certain classes of symmetry-enriched topologically ordered systems where the (local) symmetry action is either non-Abelian, or acts projectively leading to multi-dimensional local degeneracies~\cite{PhysRevB.94.224206}. We note that while non-Abelian topologically ordered systems cannot be MBL in the strict local integrability sense, they could potentially lead to fundamentally novel non-ergodic (but non-MBL) states of matter. This possibility has not been explored to date.
 
\subsection{Localization of anyonic edge modes}

The constraints on the localization discussed above also have consequences for quasi one-dimensional ``trenches'' of non-Abelian anyons $\psi$ with quantum dimension $d_\psi$. Focusing on the anyonic degrees of freedom and ignoring the parent 2D topological phase necessary to realize the anyons, we can ask whether the 1D topological phase with anyonic edge modes obtained by dimerizing the couplings can be protected to finite energy density using MBL. In the perfectly dimerized limit, the eigenstates of such a system consist of two dangling anyonic edge modes and of anyonic excitations resulting from the fusion $\psi \times \psi $ on the bulk dimerized bonds. Generalizing the discussion of non-Abelian symmetries, an MBL phase away from this perfectly dimerized limit can exist if and only if the anyons appearing when fusing $\psi$ with itself all have dimension one~\cite{PhysRevB.94.224206}. This can occur only if $d_\psi^2=p$ is an integer --- corresponding to the so-called parafermionic zero modes~\cite{1742-5468-2012-11-P11020,PhysRevB.86.195126,CAKpara,PhysRevX.4.011036,AFpara} that generalize Majorana fermions ($p=2$). While parafermions have interesting properties, especially from the perspective of topological quantum computation~\cite{RevModPhys.80.1083}, they are not enough to realize a set of universal gates. One-dimensional chains of anyons~\cite{PhysRevLett.98.160409, PhysRevLett.101.050401} whose braiding would be more interesting from a topological quantum computation viewpoint~\cite{RevModPhys.80.1083}  (such as Fibonacci anyons for example) cannot be many-body localized even by strongly dimerizing the couplings. (Note however that they can either thermalize~\cite{PhysRevB.94.235122,2017arXiv170904067C}, or form more exotic non-ergodic states of matter at strong disorder~\cite{QCGPRL}.)

\section{Concluding Remarks}
In closing, we comment on various possible future directions and subtleties not considered above. First, and most importantly, we remind the reader that  the various ``no-go'' arguments that limit the utility of MBL in protecting non-equilibrium topological orders are predicated on a specific definition of MBL. We have equated MBL with the existence of an l-bit Hamiltonian and a local tensor-product structure, with {\it all} eigenstates localized. To date, the only {\it rigorous} proofs of MBL have been given in this setting~\cite{2014arXiv1403.7837I,PhysRevLett.117.027201}. However, there are various ways in which this assumption may be too constraining. For instance, the existence of `many-body mobility edges' --- a finite energy density separating MBL and thermal eigenstates --- remains a subject of debate~\cite{PhysRevB.93.014203}. Were such an MBL mobility edge to exist, then the energy density would be an additional tuning parameter to drive transitions. Another intriguing possibility is whether thermalization can be evaded in a generic system without giving rise to an emergent l-bit description. There have been few explorations of this question~\cite{PhysRevB.94.144203,PhysRevB.95.064204,2017arXiv170904067C}; an affirmative answer would have profound implications for localization-protected order. 
  
Second, we have said nothing about the tremendous progress in understanding periodically driven (Floquet) systems. In such systems, MBL plays a subsidiary yet essential role, in preventing thermalization to an infinite-temperature state~\cite{PhysRevLett.115.030402,PhysRevLett.114.140401}. Driven MBL systems then exhibit robust Floquet phase structure~\cite{khemani_phase_2016}, characterized now in terms of eigenstates of the single-period evolution 
 operator (Floquet operator). Localization-protected order can therefore be adapted to this setting, but since energy is conserved only modulo multiples of the drive frequency, Floquet systems admit unique possibilities. These include Floquet SPTs with no equilibrium analog, and most famously, to ``time crystal'' phases whose sub-harmonic dynamical response spontaneously breaks the discrete time translational symmetry of the drive~\cite{khemani_phase_2016,pi-spin-glass,else_floquet_2016,PhysRevX.7.011026,PhysRevLett.118.030401,Zhang:2017aa,Choi:2017aa}. Intriguingly, while initially formulated for Floquet-Ising models,  time crystals   exhibit a form of spatiotemporal long-range order (glassy in space and period-doubling in time) that is {\it absolutely stable} against any perturbations that do not violate the periodicity of the drive~\cite{pi-spin-glass} --- even in the absence of any global symmetries. Recently-completed comprehensive classifications of Floquet SPTs~\cite{lindner_floquet_2011,Wang453,rechtsman_photonic_2013,PhysRevB.92.125107,von_keyserlingk_phase_2016,PhysRevB.93.201103,potter_classification_2016,PhysRevB.94.125105,PhysRevB.95.195128,2017arXiv170601888P} and detailed treatments of their symmetry-breaking properties provide elegant starting points for future studies.

A third point that we have not considered is the link between  MBL and  quantum chaos. Indeed, we have alluded to ``thermalizing'' behavior; in an isolated system we mean by this  that the Eigenstate Thermalization Hypothesis (ETH)~\cite{PhysRevA.43.2046,PhysRevE.50.888}  is satisfied. ETH is in essence a set of criteria on individual eigenstates and matrix elements between pairs of eigenstates; when these are obeyed, local observables decorrelate from their initial condition and exhibit behavior characteristic of thermal equilibrium at a temperature set by the initial (conserved) energy density. Recently, several works have begin to develop a more precise description of the onset of chaos and ergodicity, and discuss the spreading of entanglement and conserved quantities~\cite{PhysRevX.7.031016,2017arXiv170510364N,2017arXiv170508975N,2017arXiv171009835K,2017arXiv170508910V,2017arXiv171009827R,2017arXiv171206836C} with a view to exploring fundamental bounds on the dynamics of quantum information~\cite{Maldacena2016}. Progress in these directions has the potential to feed back on questions discussed here, in that they may indicate routes other than MBL whereby systems can evade thermal equilibrium, at least up to very long time scales~\cite{PhysRevLett.115.256803,PhysRevX.7.041062,PhysRevX.7.011026,1742-5468-2017-6-063105}. Whether these can in turn lead to similarly rich phase structure is an open and intriguing question.

Fourth, we have explored in detail only a simple example of topological order in the 2+1D toric code. In three dimensions, there are additional possibilities~\cite{2017arXiv171204943P} such as the so-called {\it fracton} models~\cite{PhysRevLett.94.040402,BRAVYI2011839,doi:10.1080/14786435.2011.609152,PhysRevA.83.042330,PhysRevLett.111.200501,PhysRevB.88.125122,PhysRevB.92.235136,PhysRevB.94.155128,Nussinov06102009,PhysRevB.95.245126,2017arXiv170100762V,PhysRevLett.119.257202,2017arXiv170909673P,2017arXiv171202375S} that depart significantly from the usual `emergent gauge theory' paradigm of topological order. Such models have a topological ground state degeneracy that {scales exponentially with linear system size}, and host  excitations that are partially deconfined and restricted to move only on lower-dimensional subspaces such as surfaces or lines. As the existing fracton theories are all captured by zero-correlation length Hamiltonians, the arguments given in Section~\ref{sec:LPTO} generalize directly to this new setting, so in the presence of disorder they can exhibit eigenstate fracton order (indeed, owing to the peculiarity of their excitations fracton models exhibit glassy dynamics and extremely slow equilibration even in the clean limit~\cite{PhysRevLett.94.040402,PhysRevB.95.155133}.) Quantities such as entanglement entropy~\cite{2017arXiv170509300S,2017arXiv171001744M} and the Fredenhagen-Marcu diagnostic~\cite{FractonFM}, suitably generalized, can detect fracton order in highly-excited eigenstates.

Fifth, we have provided not even a cursory discussion of the experimental complexities involved in stabilizing and exploring far-from equilibrium MBL-protected orders. Suffice to say that each of the possible venues where such behavior may arise --- such as ultracold neutral atoms or polar molecules, trapped ions, nitrogen vacancy centers in diamond, or even the solid state --- has its own advantages and possible pitfalls. 
Here we simply note that MBL may also play a role even in imperfectly isolated systems~\cite{Schreiber842,Smith:2016aa,PhysRevX.7.011034,Choi1547,2016arXiv161205249W}, as in many cases the timescales for thermalization by a bath can be extremely long~\cite{PhysRevB.90.064203,PhysRevLett.116.237203,PhysRevLett.116.160401,PhysRevB.93.094205,PhysRevLett.114.117401,ANDP:ANDP201600181}. For instance, MBL-motivated ideas are relevant to exploring the transport properties of Luttinger liquids in the spin-incoherent regime~\cite{PhysRevB.95.024201}.

Sixth, we have said little about the implications of MBL for numerical simulations. Owing to their area-law entanglement, even highly-excited states of MBL systems may be given an efficient matrix-product state representation. This suggests that techniques that exploit this property, such as the density-matrix renormalization group, may now be generalized to efficiently simulate MBL systems~\cite{PhysRevLett.116.247204,PhysRevLett.118.017201,PhysRevB.94.045111,PhysRevB.93.245129,PhysRevLett.117.160601,PhysRevX.7.021018}. Initial efforts in these directions are promising, but there is yet scope for improvement and further work.

Returning to the topics we {\it have} discussed,  the identification of localization-protected orders suggest many possible directions of further study. Perhaps most interesting is that the deconfinement often provides a route to accessing strongly-correlated states of matter that cannot be adiabatically connected to a simple free limit. For instance, $\mathbb{Z}_2$ gauge structure is a minimal ingredient in constructing simple toy models of non-Fermi liquid states in $d>1$~\cite{PhysRevB.86.045128}. With the advent of MBL and its attendant consequences for the whole eigenspectrum, it may be possible to define similar concepts 
 in an out-of-equilibrium setting. While some work has begun to explore such possibilities~\cite{PhysRevLett.119.146601}, there remain many open questions for the future, suggesting that MBL and its associated phenomena are likely to be vibrant research topics for some time to come.

\section*{Acknowledgements}
We thank A.C. Potter, A.J. Friedman, W. Berdanier, M. Kolodrubetz, S. Gazit, B.~Kang, J. Moore, S. Gopalakrishnan, R. Nandkishore, H. Ma, A.T. Schmitz, M.~Hermele,  T. Devakul, and S.L. Sondhi for past and ongoing collaborations on related topics. We are grateful to them and to D. Abanin, E. Altman, F. Burnell, A. Chandran, D. Huse, V. Khemani, W. de Roeck, F. Huveneers, W.-W. Ho, M. M\"{u}ller, P. Fendley, J. Bardarson and F. Pollmann for discussions and correspondence. We particularly thank D. Huse, V. Khemani, and R. Nandkishore for detailed and insightful comments on the manuscript. We acknowledge support from NSF Grant DMR-1455366 at the University of California, Irvine (SAP) and University of Massachusetts start-up funds (RV).
\vspace{0.1in}

\bibliography{MBL.bib}

\end{document}